\documentclass[10pt,prl,twocolumn,natbib,showkeys] {revtex4}

\usepackage{graphicx}
\usepackage{dcolumn}
\usepackage{bm}
\usepackage{amsmath}
\input{epsf}

\begin{document}

\newcommand{\note}[1]{\textbf{#1}}

\title{Fluxon readout of a superconducting qubit}
\author{\firstname{Kirill G.} \surname{Fedorov$^{1,2}$}}
\email{kirill.fedorov@kit.edu}
\author{\firstname{Anastasia V.} \surname{Shcherbakova$^{1}$}}
\author{\firstname{Michael J.} \surname{Wolf$^{3}$}}
\author{\firstname{Detlef} \surname{Beckmann$^{3}$}}
\author{\firstname{Alexey V.} \surname{Ustinov$^{1,2,4}$}}
\affiliation{ $^{1}$Physikalisches Institut and DFG-Center for Functional Nanostructures (CFN), Karlsruhe Institute of Technology, D-76131 Karlsruhe, Germany \\
$^{2}$ National University of Science and Technology MISIS,
Leninsky prosp. 4, Moscow 119049, Russia \\
$^{3}$ Institute of Nanotechnology, Karlsruhe Institute of Technology, 76021 Karlsruhe, Germany \\
$^{4}$ Russian Quantum Center, 100 Novaya St., Skolkovo, Moscow region, 143025, Russia  }

\date{\today}

\begin{abstract}
An experiment demonstrating a link between classical single-flux quantum digital logic and a superconducting quantum circuit is reported. We implement coupling between a moving Josephson vortex (fluxon) and a flux qubit by reading out of a state of the flux qubit through a frequency shift of the fluxon oscillations in an annular Josephson junction. The energy spectrum of the flux qubit is measured using this technique. The implemented hybrid scheme opens an opportunity to readout quantum states of superconducting qubits with the classical fluxon logic circuits.
\end{abstract}

\pacs{74.50.+r, 84.40.Lj, 71.15.Rf}

\keywords{Josephson junctions, fluxons, qubits}

\maketitle

Quantum computing using superconducting circuits underwent rapid development in the last decade \cite{QQ1,QQ2,SCQubits,PQ_Mart2009}. This field has propelled from quantum manipulation of single two-level systems to complex designs employing multiple coupled qubits allowing one to execute simple quantum algorithms \cite{2Q_DiCa2009,QC_Luc2012}. On the way to a practical quantum computer, a need for scalable interfaces between classical circuits and the quantum counterparts has arisen.

A quantum computer requires a set of coupled quantum bits (qubits) with control gates to manipulate them, as well as a classical computer as an interface to the quantum counterpart \cite{DiVi2000}. As for today, there are very promising candidates for the role of quantum bits from the field of solid state superconducting qubits - as phase, flux or transmon qubits \cite{QQ2,SCQubits,PQ_Mart2009,FluxQubit1,TrQ_Schr2008}. Experiments of the past decade have shown how to entangle and operate these qubits \cite{FluxQubit2,Siddiqi-2012}, and simple quantum algorithms have been demonstrated \cite{2Q_DiCa2009,QC_Luc2012}. Nowadays, a classical interface for qubits is an emerging milestone in the development of circuits with multiple solid state qubits. An efficient control and readout of several quantum bits requires a powerful classical computer in order to process the vast amount of real-time measurement data from the quantum counterpart. Researchers often use specific programmable electronic boards, e.g., field-programmable gate array (FPGA) boards, in order to meet the requirements for high processing speed, flexibility and reasonable price. However, in the near future this approach will become inefficient due to the design complexity and engineering problems of communication between many qubits and room temperature readout electronics.

Low-temperature superconducting single-flux quantum (SFQ) logic employs magnetic fluxons (Josephson vortices) in Josephson transmission lines (JTLs) as basic bits for classical computation \cite{RSFQ_Likh2012,SFQ-EU-Roadmap-2010}. Superconducting SFQ electronics offers a possibility of implementing an extremely fast, up to $500$ GHz clock speed, low-temperature classical computer \cite{RSFQ_Likh2012,SFQ-EU-Roadmap-2010}. SFQ logic can be optimized to be used at millikelvin temperatures \cite{LT-RSFQ_Inti2006,LT-RSFQ_Savi2006} and be also modified to allow for reversible operation \cite{Semenov-rev-2007,Semenov-rev-2011}. The lack of compact memory elements in SFQ electronics can be resolved by using magnetic Josephson junctions \cite{MJJ_Ryaz2012}. With the advent of superconducting quantum computing and the requirement to process a lot of data at low temperatures, SFQ electronics seems to be the natural candidate for the role of an interface between room temperature electronics and its quantum counterpart. In this Letter, we experimentally demonstrate the key link between a superconducting flux qubit and SFQ-based electronic circuit.

\begin{figure}[h!]
\begin{center}
\resizebox{0.9\columnwidth}{!}{
\includegraphics{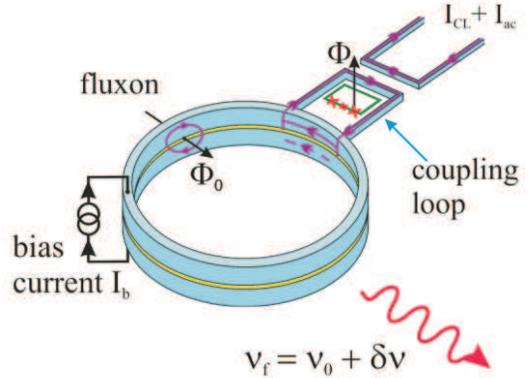}}
{\caption{An annular Josephson junction coupled to a flux qubit. The junction is homogeneously biased by an external current source $I_b$. The flux qubit is controlled by a dc control line current $I_{CL}$ and a microwave signal $I_{ac}$ at the frequency of $\nu_e$. The purple lines schematically show the flow of the electrical currents. The fluxon equilibrium oscillation frequency $\nu_0$ is shifted by an amount $\delta\nu$ due to fluxon scattering on the current dipole produced by a flux qubit. Microwave radiation from the fluxon is picked up by a capacitively coupled microstrip antenna and later fed to a cryogenic amplifier (not shown).} \label{AJJ+Qubit}}
\end{center}
\end{figure}
A single fluxon (alias Josephson vortex) in an underdamped JTL has properties of a relativistic particle carrying a magnetic flux quantum $\Phi_0=h/2e$ \cite{Fulton}. The size of a vortex can vary from a few to several hundreds of microns, depending on the critical current density $j_c$ and vortex velocity in JTL. By applying a bias current, the vortex can be accelerated up to the Swihart velocity $c_S$, which is the characteristic velocity of electromagnetic waves in JTL. The dynamical properties of the fluxon resemble a classical particle with a well-defined mass and velocity. In SFQ technology, fluxons are used to realize logic operations by transmitting and detecting them in JTLs.

The idea of using fluxons for detecting the state of flux qubits was initially proposed and theoretically analyzed in Ref.~\cite{FR-Averin}. A ballistic fluxon moving in a JTL that is weakly coupled to a superconducting flux qubit can be used to readout the quantum state of the qubit \cite{FR-Averin,FR-Herr}. There is a potential advantage of the fluxon readout over the usual dispersive readout of qubits with the use of microwave resonators, usually referred to as QED readout \cite{Wallraff-Nat-04}. The distribution of the electro-magnetic field in the usual bi-coplanar resonator is such that it perturbs the qubit as long as there are photons in the resonator. For the fluxon readout, the back-action on the qubit occurs only for a very short time, when the fluxon passes by the qubit. This time depends on JTL parameters with the typical time scales of about a few picoseconds. This short-time interaction improves isolation of the qubit from the environment and hence makes it possible to increase the qubit coherence.

For our experiments, we employ a JTL formed by a continuous annular Josephson junction (AJJ). JTLs of such closed topology (see Fig.~\ref{AJJ+Qubit}) are used as on-chip SFQ clocks \cite{Vernik-clock} and their most significant advantage is the conservation of the total magnetic flux initially trapped in JTL. One can create a fluxon on demand in a flux-free JTL by applying a current through a pair of injectors \cite{FluxonInsert}. To couple a flux qubit to the fluxon inside the AJJ, it is necessary to engineer the interaction between two orthogonal magnetic dipoles. To facilitate this interaction, we have added a superconducting coupling loop embracing the flux qubit, as shown in Fig.~\ref{AJJ+Qubit}. The current induced in the coupling loop attached to the AJJ is proportional to the persistent current in the flux qubit. Thus, the persistent current $I_p$ in the qubit manifests itself in the AJJ as a current dipole with an amplitude $\mu = k I_p / (j_c \lambda_J W)$ on top of the homogeneous background of bias current $\gamma = I_b / I_c$. Here, the coefficient $k \sim R_D \cdot M$ reflects the inductive coupling $M$ between the qubit and the coupling loop as well as the fluxon differential resistance $R_D = \frac{\partial \nu_f}{\partial I_b}$, $W$ is the width of the JTL, $\lambda_J$ is the Josephson penetration depth, $j_c$ is the critical current density, $I_b$ is the bias current, and $I_c$ denotes the critical current of the AJJ. A theoretical description of interaction between a single Josephson vortex and a current dipole in the AJJ can be given by the perturbed sine-Gordon equation (PSGE) \cite{MalUst2004}
\begin{equation} \frac{\partial^2\varphi}{\partial
t^2}+\alpha\frac{\partial\varphi} {\partial
t}-\frac{\partial^2\varphi}{\partial x^2}=\gamma-\sin
(\varphi)+\mu(\delta(x-d/2)-\delta(x+d/2)) \label{PSGE}
\end{equation}
with the corresponding boundary conditions
\begin{equation}
\varphi(-l/2,t) = \varphi(l/2,t) + 2\pi n;
\frac{\partial\varphi(-l/2,t)}{\partial x}=
\frac{\partial\varphi(l/2,t)}{\partial x}, \label{Bndr}
\end{equation}
where $n$ is the number of trapped fluxons, $\alpha = \omega_p / \omega_c$ is the damping parameter, $l = L / \lambda_J$ is the normalized junction circumference, $d = D / \lambda_J$ is the normalized dipole length, $\omega_p = \sqrt{2 e I_c / \hbar C}$ is the plasma frequency, $\omega_c = 2 e I_c R_N / \hbar$ is the characteristic frequency, $C$ is the AJJ capacitance, $R_N$ is its normal state resistance.

Under the influence of the bias current $I_b$, the fluxon rotates in the AJJ. This rotation produces both dc and ac voltage components, with the magnitude of the former proportional to the frequency of the latter. In our detection scheme, we directly measure the ac voltage induced by the fluxon as microwave radiation using a low-noise cryogenic microwave amplifier \cite{FR-Fedorov}. Our approach of acquiring the fluxon radiation frequency provides a much greater precision than dc voltage measurements.

The studied JTL was fabricated by using photolithography and the standard Nb/AlO$_x$/Nb trilayer process with critical current density $j_c \simeq 800$ A/cm$^2$. The estimated Josephson penetration depth is $\lambda_J \simeq 13\, \mu$m, the Josephson plasma frequency is $\omega_p/2\pi \simeq 154$ GHz, and the estimated damping parameter at the operating temperature is estimated as $\alpha \simeq 0.02$. The circumference of the junction $L = 880~\mu$m determines the maximum frequency of the radiation at about $15$ GHz, corresponding to the fluxon moving with the Swihart velocity $c_S$. The width of the AJJ was $W = 5 \mu$m, the dipole length $D = 35 ~\mu$m and its fluxon-free critical current $I_c = 35$ mA. The flux qubit was deposited using the standard aluminum shadow evaporation process \cite{FluxQubit1} on top of the pre-fabricated niobium structures. Designed fabrication parameters for the Josephson junctions in the flux qubit loop were the following: critical current $I_c = 500$ nA, alpha factor $\alpha_q = 0.63$, ratio of charging and Josephson  energies $E_C / E_J = 0.0034$.
\begin{figure}
        \begin{center}
        \includegraphics[width=\linewidth,angle=0,clip]{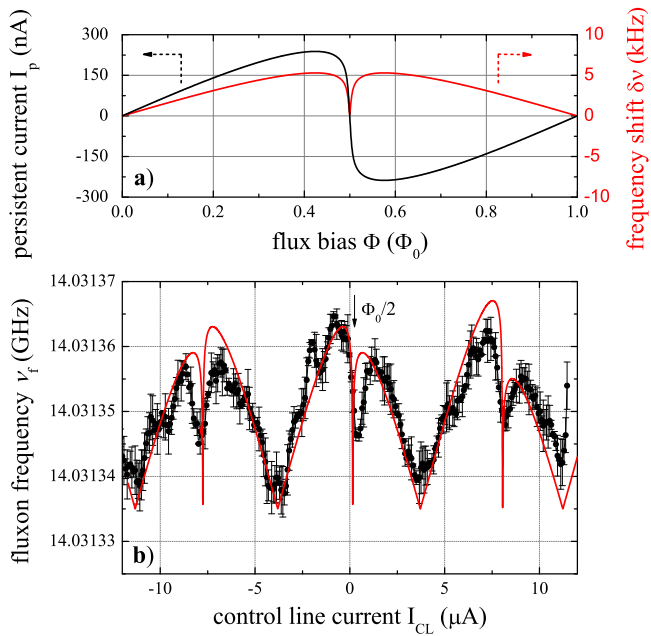}
        \end{center}
    \caption{ \textbf{ a) } The theoretical persistent current $I_p$ of the ground state of the flux qubit versus magnetic frustration (black line) calculated for our flux qubit parameters. The red line shows the expected fluxon frequency shift in kHz for $\gamma = 0.4$ and $\mu = 0.05$. \textbf{b)} The experimentally measured modulation of the fluxon oscillation frequency due to the coupling to the flux qubit. Black dots show the measured mean frequency of fluxon oscillations. Every point consists of $100$ averages; bias is $\gamma = 0.39$. The red line shows the corresponding fit.}
    \label{FreqMod}
\end{figure}

To experimentally test the qubit readout scheme, the temperature was stabilized at $T = 70$ mK, well below the superconducting transition temperature $T_c$ of aluminum forming the qubit. The AJJ was biased at a fixed current $I_b$. Then, we varied the current through the control line $I_{CL}$ in order to change the magnetic flux through the flux qubit. Due to periodic variations of the persistent current $I_p$ in the qubit loop versus the control line current $I_{CL}$, the current dipole strength $\mu$ is also changing periodically. The persistent current of the flux qubit in the ground state was calculated numerically \cite{OrlMoo_PRB99} and is depicted by the black solid line in Fig.~\ref{FreqMod}.a.

When a ballistic fluxon scatters on a positive current dipole, it first gets accelerated and then decelerated by the dipole poles. In the case of absence of damping and bias current, the change of its circulation frequency is determined only by the dipole polarity and the dipole amplitude. In the presence of finite damping $\alpha$ and homogeneous bias current $\gamma$, the total propagation time becomes dependent on the complex interplay between bias current, current dipole strength, and damping. Our previous numerical simulation of Eq.~(\ref{PSGE}) and perturbation analysis \cite{FR-Fedorov} have shown that, for fluxon velocities close to $c_S$, the shift of fluxon oscillation frequency almost does not depend on the sign of the dipole $\mu$ and depends only on the absolute amplitude of the dipole $\mu$ and the bias current $\gamma$, looking like it is shown by the red line in Fig.~\ref{FreqMod}.a (where the black curve $I_p$ was taken as the current dipole strength $\sim \mu$). As the dipole length $d \simeq 3$ is much larger than the characteristic size of the fluxon at relativistic velocities, the contributions of the separate dipole poles to $\delta\nu$ is additive and is not dependent on their order. For smaller bias currents (small fluxon velocities) and smaller dipole lengths, when $d$ is comparable with the fluxon size, $\delta\nu$ becomes dependent on the dipole sign.

The back-action of the propagating fluxon on the flux qubit can be estimated from the magnetic flux $\Phi_{ba}$  which the fluxon excites in the qubit loop, depending on the geometrical inductances of the loop and the qubit. For our test sample, the estimated maximum back-action flux $\Phi_{ba}$ is about $35$ m$\Phi_0$. Considering that the fluxon interacts with the qubit on the time scale of $\tau \simeq 3$ ps, we can calculate the qubit phase shift $\delta\phi = \int_0^{\tau} \frac{E_{12}(t) - E_0}{\hbar} dt$ induced by the fluxon. We find that $\delta\phi \sim 2 \cdot 10^{-3} \cdot 2\pi$ per fluxon revolution in the AJJ. It corresponds to the dephasing time due to the readout back-action of $T_{ba} \sim 35$ ns which is consistent with the observed linewidth of the flux qubit spectra. This short coherence time may be considered as a strong disadvantage of the fluxon readout. However, simply by decreasing the critical current density to $j_c \sim 10$ A/cm$^2$ and, at the same time, by increasing the length of the coupling loop, one should be able to decrease the back-action flux $\Phi_{ba}$ and increase $T_{ba}$ by, at least, 2-3 orders of magnitude while keeping the current dipole amplitude $\mu$ at the same level.
\begin{figure}
        \begin{center}
        \includegraphics[width=\linewidth,angle=0,clip]{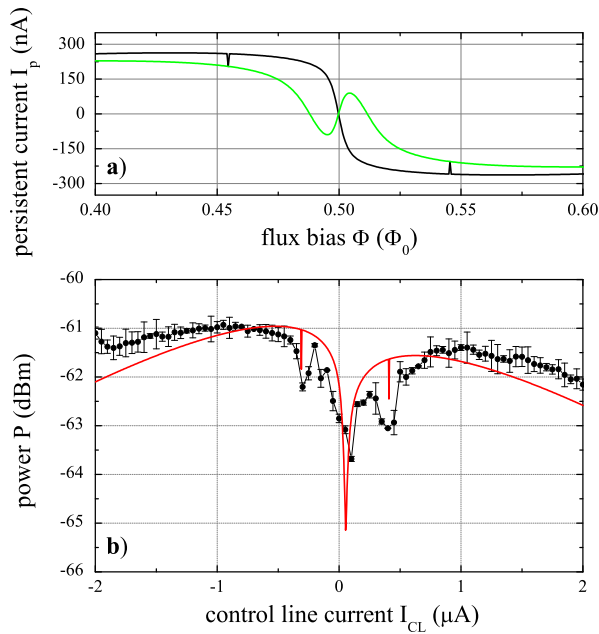}
        \end{center}
    \caption{ \textbf{a)} The theoretically expected persistent current $I_p$ for the ground state of the flux qubit versus magnetic flux (black line). The spikes indicate the presumed transitions to the first excited state at flux bias points corresponding to $E_{12} = 14$ GHz flux qubit energy splitting. The green line shows the expected persistent current $I_p$ for the first excited state. \textbf{b)} Modulation of the fluxon's oscillation frequency due to the coupling to the qubit, measured in the power domain. Measured power $P$ relates to the power at the fixed frequency offset $+50$ kHz from the fluxon mean oscillation frequency $\nu_f$. Every point consists of $10$ averages with video filter bandwidth of $10$ Hz. The red line depicts the corresponding theory fit.}
    \label{AmpMod}
\end{figure}

Figure~\ref{FreqMod}.b shows the experimental shift of the fluxon oscillation frequency due to coupling to the flux qubit at the fixed bias current $\gamma = 0.39$. We measured a Lorentzian radiation peak of fluxon oscillations in the AJJ for fixed values of $I_{CL}$ and then determined the mean frequency $\nu_f$ for which the radiation power was maximal. To fit our experimental data to the theory, we take into account the parasitic coupling between the control line current and the fluxon leading to an additional linear shift of $\nu_f$ versus $I_{CL}$. The corresponding fit is presented by the red line in Fig.~\ref{FreqMod}.b showing fairly good correlation between theory and experiment. Smaller irregular peaks visible in Fig.~\ref{FreqMod}.b are low-frequency fluctuations, most probably arising from trapped Abrikosov vortices in the superconducting leads of the AJJ. Presumably, these Abrikosov vortices are also responsible for the parasitic flux offset in Fig.~\ref{FreqMod}.b as this offset varies for different cooldowns.
\begin{figure}
        \begin{center}
        \includegraphics[width=\linewidth,angle=0,clip]{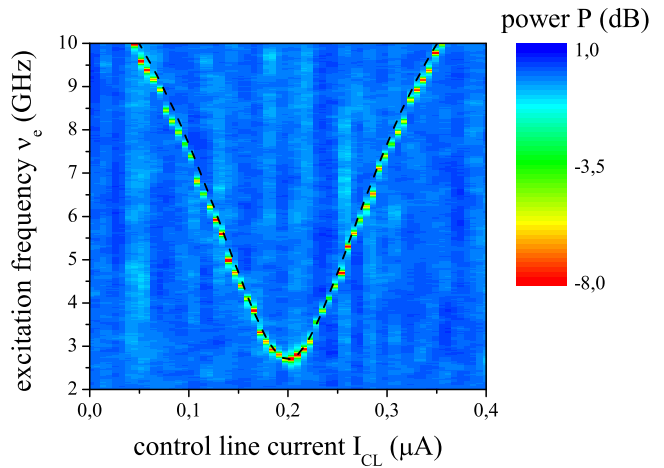}
        \end{center}
    \caption{Fluxon response on the microwave excitation applied to the flux qubit. Color scale corresponds to the detected power. Measured power $P$ relates to the power at the fixed frequency offset $+50$ kHz from the fluxon mean oscillation frequency $\nu_f$. Every point consists of $10$ averages with video filter bandwidth of $1$ Hz. The flux qubit spectrum can be clearly seen as the red-yellow curved trace. The black dashed line depicts the corresponding fit of the flux qubit.}
    \label{Spectrum}
\end{figure}

To speed up the measurement process and decrease noise, we switched to zero-frequency span and measured the power at the fixed frequency $\nu_f + 50$ kHz. Figure~\ref{AmpMod}.b shows the results of such measurements around $\Phi \sim \Phi_0/2$. A sharp dip is clearly visible at the $\Phi_0/2$ flux as well as two other smaller symmetric satellites around it. These satellites occur when the energy splitting $E_{01}$ between the ground and excited states of the qubit coincides with the fluxon oscillation frequency $\nu_f$. This leads to the change of persistent current in the qubit for two values of flux bias $\Phi$ shown by the black line in Fig.~\ref{AmpMod}.a. The theoretical fit of the data for the fluxon resonance frequency $\nu_f = 14.031$ GHz is presented by the red line in Fig.~\ref{AmpMod}.b. At resonance, pumping at the fluxon frequency should lead to Rabi oscillations in the flux qubit between its ground and excited states and, therefore, the measured signal reflects a mixture of the ground and the first excited states. This operation is similar to resonant interaction between qubit and resonator in a cavity QED setup.

As the last step, we performed microwave spectroscopy of the qubit. We applied an additional continuous excitation tone from an external microwave source at the frequency $\nu_e$ and swept the control line current between two resonant dips (see Fig.~\ref{AmpMod}.b). The signal of the fluxon readout for every flux bias point without microwaves was subtracted from the actual response with microwaves to get rid of an unwanted background slope. The resulting color plot of $P(\nu_e,I_{CL})$ is shown in Fig.~\ref{Spectrum}. One can clearly recognize the hyperbola of the flux qubit spectrum \cite{FluxQubit1} as a white-blue curved line between $2$ and $10$ GHz, with the minimal energy splitting $\Delta \simeq 2.7$ GHz. We can very well fit the measured spectrum by the theoretical curve (shown by the black dashed line in Fig.~\ref{Spectrum}) for the following parameters: critical current $I_c = 320$ nA, alpha factor $\alpha_q = 0.58$, ratio of charging and Josephson energies $E_C / E_J = 0.0034$. It is worth noting that the fluxon readout in this experiment was operated in the classical rather than quantum regime. Employing fluxons to interact with or act as qubits in the quantum regime, as discussed theoretically \cite{Kemp2002} and also detected experimentally \cite{Wall2003,Pric2010}, would be very exciting and challenging in the future.

In conclusion, we have demonstrated a coupling of an oscillating single fluxon in an annular Josephson junction to a flux qubit. The persistent current in the flux qubit has been detected as a shift of the fluxon oscillation frequency. Resolution of the measurement scheme is high enough to measure transitions between quantum states as shown by the acquired energy spectrum of the flux qubit. The sensitivity of this readout can be significantly improved in various ways, for instance, by lowering the critical current density $j_c$ or decreasing the width $W$ of the long Josephson junction, as these both increase a current dipole strength $\mu$ and therefore a fluxon frequency response. Our results prove the possibility of detecting quantum states by fluxon readout and, thus, open the way to applying SFQ technique in the field of superconducting quantum computing.

The authors would like to acknowledge stimulating discussions with M. Jerger and J. Lisenfeld. This work was supported in part by the EU project SOLID, the Deutsche Forschungsgemeinschaft (DFG) and the State of Baden-W{\"u}rttemberg through the DFG Center for Functional Nanostructures (CFN).

\end{document}